\documentclass{elsart}

\usepackage{natbib}
\usepackage[T2A]{fontenc}
\usepackage{amssymb}

\usepackage{color}
\usepackage{graphicx}
\usepackage{amsmath}
\usepackage{amsfonts}

\newcommand{\Lb}{L}
\newcommand{\mycap}{H}
\newcommand{\myc}{{\sf c}}

\newcommand{\mye}{{\sf e}}
\newcommand{\myf}{{\sf f}}
\newcommand{\myg}{{\sf g}}
\newcommand{\myj}{{\sf j}}
\newcommand{\myh}{{\sf h}}
\newcommand{\mym}{{\sf m}}
\newcommand{\myn}{{\sf n}}
\newcommand{\myo}{{\sf o}}
\newcommand{\myp}{{\sf p}}
\newcommand{\myt}{{\sf t}}
\newcommand{\mya}{\sf{a}}
\newcommand{\myx}{{\sf x}}
\newcommand{\myy}{{\sf y}}

\newtheorem{lemma}{Lemma}

\begin{document}
\begin{frontmatter}

\title{}
\title{Existence of a polyhedron which does not have a non-overlapping
pseudo-edge unfolding \thanksref{label1}}

\thanks[label1]{Work is supported by RFBR  
grants 08-01-00565, 08-01-91202, DFG Research Unit ``Polyhedral Surfaces''}

\author{Alexey S Tarasov }

\address{ISA RAS, Moscow, Russia, }

\begin{abstract}

There exists a surface of a convex polyhedron $P$ and a partition
$L$ of $P$ into intrinsically flat and convex geodesic 
polygons such that there are no
connected "edge" unfoldings of $P$ without self-intersections
(whose spanning tree is a subset of the edge skeleton of $L$).

\end{abstract}


\end{frontmatter}

\section{Introduction}

Let $S$ be an abstract $2$-dimensional polyhedral surface with
$n$ vertices. 
We say that $S$ is {\it intrinsically convex } if the curvatures
$\omega_1,\omega_2,\ldots,\omega_n$ of all vertices satisfy 
$0 < \omega_i < 2\pi$ for all $i \in [n]$. Of course the surface
of every convex polytope in $\mathbb R^3$ is intrinsically convex.
The following result shows that there are essentially no other
examples:

{\it (Alexandrov's existence theorem). Every intrinsically convex
$2$-dimensional 
polyhedral surface homeomorphic to a sphere is isometric to the
surface of a convex 
polytope $P \in \mathbb R^3$ or to a doubly covered polygon.
}

There is a conjecture: {\it every convex polyhedron has an edge
net}. This question has a long story. 
The earliest reference to edge unfolding was made by Albreht
D\"urer [2].
At first time this conjecutre was explicilty posted by Sheppard in 1975 \cite{She75}.

At first glance the conjecture seems to be true. 
If we take a polyhedron, it is very easy to find its unfolding.

Even to find a polyhedron with at least one 
overlapping unfolding is not trivial.

\begin{figure}[htb]
\begin{center}
\begin{tabular}{c c}
\includegraphics[clip]{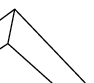}  &
 \includegraphics[clip]{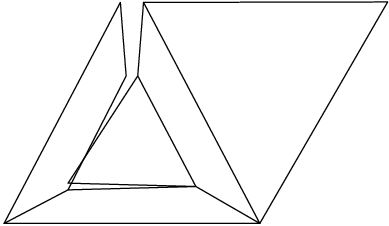} \\ 
\end{tabular}
\end{center}
\caption{Overlapping unfolding.}
\label{unfolding}
\end{figure}

On fig.~\ref{unfolding} we can see such example of overlapping unfolding 
to the thin truncated pyramid.
It is clear that the unfolding has overlapping because 
the curvature (angle defect) of the upper base is very small and
the left angle of the upper base is noticeable less than $\pi/2$. 
We will use such idea later.

It is convenient to consider this conjecture by thinking about 
a polyhedral surface as a 
metric space without embedding into $\mathbb R^3$. 

According to Alexandrov's theorem, such embedding always 
exists and we do not need to care about it.

It is easy to find vertices on a surface -- they have 
total sum of angles less than $2\pi$. But points on edges and
on faces do not have such distinctive features. 

There is no simple way without a lot of calculations to find
edges skeleton.
Alexandrov's theorem says that such skeleton exists and unique,
but does not give any way how to find it.
There is a more constructive proof of this theorem (\cite{BobIzm06}).
But still this proof need make a lot of calculations to find
edges.

Then we have the following question: does edge-skeleton a really
important
in the D\"urer conjecture.

We can formulate a similar conjecture, which does not depend from
edge-skeleton:
{\it
Let $P$ be a polyhedral surface homeomorphic to a sphere. 
Let $L$ be a partition of  $P$ into geodesic polygons, 
flat and convex in terms
of the intrinsic metric (see def. 3) 
Does there exist a non-overlapping unfolding of this
polyhedral surface, whose spanning tree lies inside edge skeleton
$E(L)$?
}

In this paper we construct a counterexample to this question.

A similar question about geodesic triangulated 1-skeletons was
asked by Jeff Erickson~\cite{Eric06}. 
{\it Let $T$ be an arbitray triangulation of a 
convex polytope whose vertices are the vertices of 
the polytope and whose edges are geodesics.
Can the surface by unfolded without 
self-overlap by cutting it along edges of $T$?
}

It appears that for Erickson's question similar counterexample
can also be constructed.

This question was inspired by another very 
interesting instinsic conjecture on the same conference:

Bobenko's conjecture~\cite{Bob06}
{\it Can the boundary of a convex polytope be unfolded 
into the plane without self-overlap by cutting the 
surface along of the Delaunay triangulation $T$ 
of the boundary? ($T$ can have loops and multiple edges, 
but the faces are triangles).
In particular, if all faces of a polytope are acute triangles, 
can it be unfolded by cutting along the edges of the polytope? }

Here another two similar conjectures by O'Rourke~\cite{ORED}:
{\it 
Prove (or disprove) that every convex polyhedron all of whose
faces have no acute angles
(i.e., all angles are $\ge 90^\circ$)
has a non-overlapping edge-unfolding
}

{\it  
Prove (or disprove) that every triangulated convex polyhedron,
all of whose angles are non-obtuse, i.e., $\le 90^\circ$,
has a non-overlapping edge-unfolding.
If the polyhedron has $F$ faces, what is the fewest nets
into which it may be cut along edges? 
}

Outline of the proof: \\
\begin{enumerate}
\item
Criteria for non-existence of an unfolding for a special polyhedral
surface with border. 

Here we will give a sufficient condition for non-existence
of a one-piece unfolding which is generalizing overlapping 
as on fig.~\ref{unfolding}. 
This condition is defined on the plane 
in terms of the weighted center of subtree vertices with respect
to their
corresponding root edges.

\item Construction a spiral-like gadget.

Here we will show the  polygon $T$ and the partition $L_T$ with
some special properties. In
particular, the polyhedral surface in some sense close enough to this partition does not have an unfolding. 
does
not have an unfoldable net.

\item Construction of a counterexample.

Here we will show a counterexample (in which additional vertices
with no
curvature are used).

\item Proof that the counterexample does not have an unfolding.
Construction of a counterexample
 without additional zero-curvature vertices.

\item A simpler counterexample.

\end{enumerate}

\section{Definitions}

{\bf Definition 1.}
An unfolding is a set of polygons with a rule for gluing the
boundaries.

{\bf Definition 2.} The curvature $curv(v)$ of a vertex $v$ is
the
difference between $2\pi$ and the sum of planar angles incident
 to this vertex.

{\bf Definition 3.} 
A partition $L$ of a surface of a polyhedron is 
called  {\it convex geodesic } 
if the following conditions hold:\\ 
1. Any vertex with non-zero curvature is a vertex of $L$;\\
2. Any part $P_i$ of the partition $L$ unfolds into a flat convex polygon.

{\bf Definition 4.} 
{\it $L$-unfolding } is a unfolding of $P$  
spanning tree of which is a subset of edge-skeleton $E(L)$ of the convex
geodesic partition $L$.

{\bf Definition 5.} An {\it infinitesimally curved polyhedron}
is
a convex polygon $P$ on the plane $T$ together with
finitely many internal points $p_i \in int(P)$ in general position
and weights $\alpha_i$ for each internal point ($\alpha_i \ge
0, \sum_{\forall i} \alpha_i=1$);
Vertices $q_i$ are vertices of polygon $P$.

Let us assume that diameter  $max(x,y) \forall x,y \in P$ 
of $P$ is equal to $1$.

{\bf Definition 6.} For a given infinitesimally curved 
surface $P$ 
and $0 < \beta < 2 \pi$ a 
{\it cap} $\mycap$  is a polyhedral surface with border,
if the  
following conditions hold: \\
1. $\mycap$ above plane $T$ and curved downward. \\
2. The projection of $\mycap$ on the plane $T$ is $P$. \\
3. Points $p_i$ are projections of vertices $h_i$ of $\mycap$
on the plane $T$. \\
4. Curvature $curv(h_i) = \alpha_i \cdot \beta$; \\

The existence
of a cap follows this theorem:

 {\it Alexandrov's theorem (\cite{Alex49}, Theorem 21, Part.
II \S 5, p. 118) 
For any set of points $A_1, \ldots, A_m$ on the
plane $T$ and any set of numbers $\omega_1, \ldots, \omega_m$,
such that: \\ 
1. $0< \omega_i < 2\pi $ for all $1\le i \le m$\\
2. $\sum_{i=1}^m \omega_i = 2 \pi$, \\
there exists
an unique infinite convex polyhedral surface 
 $Q$ with the following properties: \\
1. Points $A_1, \ldots, A_m$ are projections of the vertices
$h_i$ of $Q$; \\
2. Curvature $curv(h_i)=\omega_i$.}\\

Put to any point $q_i$ curvature 
$(2 \pi - \angle q_i)  (1 - \frac {\beta} {2 \pi})$, 
to point $p_i$ curvature $\alpha_i \beta$.
Total sum of curvature is $2 \pi$.
By Alexandrov's theorem there exists an unique infinite polyhedral surface
$Q$ satisfying these conditions.
Suppose $\mycap$ is a part of $Q$ and $\mycap$ projects
into $P$.
The $\mycap$ is a required cap. 

Note that  $\mycap$ depends continuously on $\beta$.
Indeed, the map of $P$ into a set of points $A_i$ and numbers $\omega_i$
is continuous. 
Thus, the reverse map not only unique but also continuous.


{\bf Definition 7.} Let $L$ be any partition of the polygon
$P$ into convex polygons $P_i$, such that the vertices of $L$
are the vertices of $P$ and points $p_i$.
Denote by $E(L)$ the 1-skeleton of the partition $L$.

Note that the partition $L$ does not have 
any correspondence with faces of $\mycap$.

{\bf Definition 8.} 
Define a map $f: P \to \mycap\subset \mathbb R^3$. 
Let triangulation $L'$ be some subdivided partition $L$ 
if the  following conditions holds:\\
\begin{enumerate}
\item
 For every vertex $p_i$ and $q_i$ the projection 
 of $f(x)$ into $T$ is $x$. 
 \item  For every edge $xy$ of the triangulation $L'$, an image 
$f(xy)$ is a shortest path between $f(x)$ and $f(y)$ stretched
evenly. 

Hence points $p_i$ in general position, for small enough $\beta$
edges of $f(L')$ in $\mycap$ are defined unique and 
continuously depending from $\beta$. 

\item For a triangle $\triangle xyz$ of $L'$ we can unfold 
the corresponding geodesic triangle $\triangle f(x) f(y) f(z)$ 
into a plane. Then $f$ is the unique affine transformation
of $\triangle xyz$ to $\triangle f(x)f(y)f(z)$. 

\end{enumerate}
It is clear that this map $f$ is one-one.

Denote by $h_i=f(p_i)$ a vertices of the hat $H$. 

When it is clear from context we will denote 
an objects (partition and cuts) and its image similarly.

{\bf Definition 9.} A subgraph $G \subset E(L) \setminus \partial
P$ 
is called
a {\it cut} if the following conditions hold:\\
(i) $G$  contains all points with positive weights; \\
(ii) $G \cup \partial P$ is connected.\\
(iii) $G$ does not contain cycles.

{\bf Definition 10.} Denote by $U_G$ 
the unfolding of $\mycap$ 
corresponding to the cut $G_{\beta}$. 

We consider that the number of components of $U_G$ can be more than
$1$ 
This number is equal to the number 
of independent cycles of the graph $G \cup \partial P$.  

{\bf Definition 11.}
For a given cut (of graph) 
$G$ on the polyhedral surface $\mycap$ let 
$g:\mycap->U_G$ be a natural map. 
If the point $x \notin G$ then $g(x)$ is a single point.

\begin{figure}[htb]
\begin{center}
\includegraphics[clip]{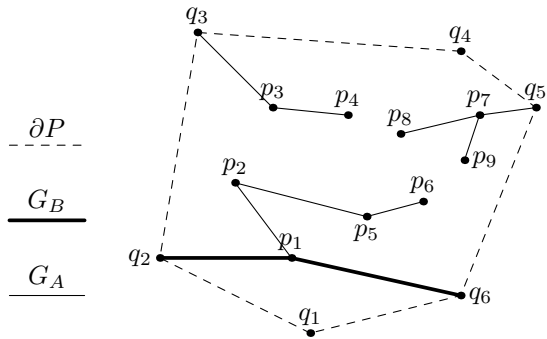} 
\end{center}
\caption{Graphs $G_A$ and $G_B$}
\label{gab}
\end{figure}

{\bf Definition 12.} An edge $e$ of a cut $G$ is
called an { \it $A$-edge}
if $e$ disconnects one of its vertices from $\partial P$.
Otherwise it is called a {\it $B$-edge} (fig. \ref{gab}).

Note that if a graph $G \cup \partial P$ contains 
only $2$ faces (inside and outside $\partial P$),
then  $G$ does not contain
$B$-edges.

Denote by $G_A$ the graph consisting of the $A-$edges of the
cut $G$.
Define graph $G_B$ similarly.

Note several facts about $G_A$ and $G_B$:
Each component of $G_A$ is a tree, which has one common point
(outfall) with $\partial P$ or $G_B$.
All endpoints of $G_B$ lie in $\partial P$.

{\bf Definition 13.} Given a cut $G$, we call a vertex $p_1$ 
 {\it upstream }
to a point $p_2$ ($p_1 >p_2$) if any path connecting $p_1$
with $\partial P$ passes through $p_2$. Similarly, we define
the upstream partial order on $A$-edges. 
Similarly we can define a {\it downstream} one.

From every vertex $v$ in $G_A$ goes only one downstream edge.

{\bf Definition 14.} Consider a neighborhood $B(y,r)$ of a point
$y$  on some $A$-edge $e$ ($y$ is not a vertex)
where $r$ is small enough 
that $B(y,r) \cap G$ is a part  of the edge $e$.

\begin{figure}[htb]
\begin{center}
\includegraphics[clip]{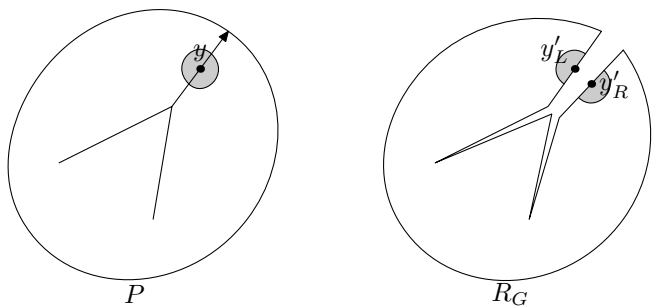} 
\end{center}
\caption{Left and right images of point $y$.}
\label{lr}
\end{figure}

Image of $B(y,r)$ in the unfolding $U_G$ consists of 
two components (fig.~\ref{lr}).
Each of these parts lies on one side of $e$ -- left or right
(we are looking downstream $e$).
Denote these parts respectively by $B_L(y,r)$ and $B_R(y,r)$.

Denote by $y'_R$ ($y'_L$) an image of $y$ 
laying in $B'_R(y,r)$  ($B'_L(y,r)$).

Similarly, we can define left and right images of any $A$-edge.

Vertex $p_i$ can have more than two images.
If $p_i$ has an incident downstream $A-$edge $e$, 
two images of $p_i$ are $u_{i,L}$ and $u_{i,R}$ 
corresponding to the images $e_L$ and $e_R$.

\begin{figure}[htb]
\begin{center}
\includegraphics[clip]{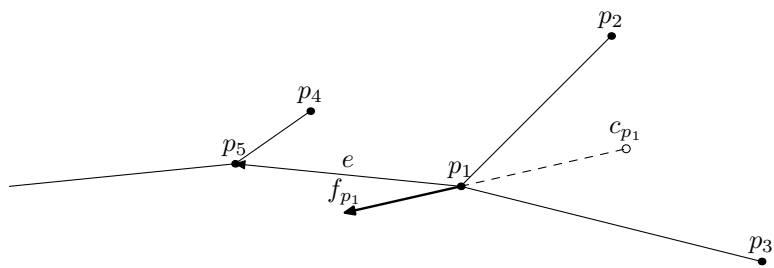} 
\end{center}
\caption{Stream for point $p_1$}
\label{stream2}
\end{figure}

{\bf Definition 15.} A {\it stream } of any point $x$ on an $A$-edge
(including the vertices) is the
vector $\sum_{p_i > x} \alpha_i (x- p_i x)$. Another way to define
the stream is: 
A {\it rotation center} $c_x$ is $\alpha_x^{-1} \sum_{p_i
> x} \alpha_i p_i$ (the weighted barycenter).
Then the stream
is equal to
$$f_x = \alpha_x  (x - c_x), \eqno{(1)} $$
where $\alpha_x = \sum_{p_i > x} \alpha_i$ (fig. \ref{stream2}).

Term {\it stream} is called by analogy with  a river.
I imagine that graph $G$ is a river net. The stream is 
the  force and the direction there river wants to flow.
Later we will show that for a non-overlapping unfolding the 
angle between the direction 
of the stream and the river (edges) should be no more than $\pi/2$.

{\bf Definition 16.} A cut $G$ is called {\it admissible 
} 
if for every point $x in G_A$ with $\alpha_x>0$ 
$$<x- c_x, e>/|e| \ge 0. \eqno(2)$$

We assume that the edge $e$ is directed from the upper
to the lower vertex.

If $x$ is a vertex of $G$, the condition $(2)$ defines a 
half-plane where edge $e$ should lie.

Denote by $l(G)$ the maximum of $- <x-c_x,e>/|e|$.
For a non-admissible cut $G$ $l(G)>0$.

Note that if we draw away from the cut $G$ all edges $e$
with $\alpha_e=0$, the obtained graph $G'$ also is a cut
without changing its admissibility.

{\bf Definition 17.} 
Denote by $l(L) = min_{\forall cut G, G_B=\emptyset} l(G)$.

A partition $L$ is called {\it non-admissible } if 
any admissible cut $G$ of $L$ contains non empty $G_B$. 

As $L$ has finitely many vertices, it has finitely many subgraphs,
Thus, the number of possible cuts $G$ with $G_B =\emptyset$  
is also finite.
Hence, for a non-admissible partition $L$ $l(L)>0$.

\section{Criteria for non-existence of an unfolding
in $\mycap$} 

In this section we will show, that
where is no $L$-unfolding for the polyhedral surface $\mycap$ 
to the infinitesimal surface with the non-admissible partition $L$ 
when curvature $\beta$ of $mycap$ is small enough.

Unfolding is one or several polygons. 
We can arrange

\begin{figure}[htb]
\begin{center}
\includegraphics[clip]{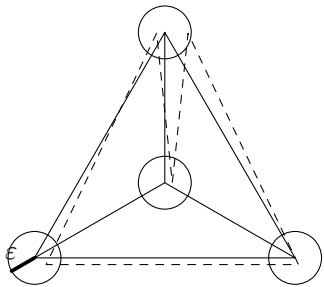} 
\end{center}
\caption{Arrangement of $P$ and 
corresponding $U$.}
\label{arrangement}
\end{figure}

Let us try to arrange these polygons on the plane $T$ 
to make any point $y \in U_G$ be as close 
as possible to its pre-image (fig.~\ref{arrangement}).

Then $\beta=0$, $H$ is a flat polygon equal to $P$.
And unfolding $U_G$ is equal to $P$.

Since $H$ depends continuously on $\beta$
for a small enough $\beta$ we can arrange $P$ and $U_G$ on 
the same plane in such way that:\\ 
1. The distance $$|x x'| < \varepsilon, \eqno(3.1)$$ 
for any point $x \in P$ and any image of it $x' \in U_G$  \\
2. For any segment $xy \in P$ and its 
connected image 
$x'y'$,  $$\left|\, |x-y| - |x'-y'|\, \right| 
< \varepsilon |x-y| \eqno(3.2).$$

Let $U_G$ be an unfolding combined with a polygon $P$
satisfying (3.1) and (3.2).
Consider some point $x$  on the cut $G$ and its two images $x'_L, x'_R$.
Denote by $\tilde c$ a point satisfying 
$$ x'_R - x'_L = \beta \alpha_x M_{rot} \cdot (x - \tilde  c_x) 
, \eqno(4)$$
where $M_{rot}$ is a $\pi/2$-clockwise rotation matrix.

\begin{lemma}

For a small enough $\beta$ 
and  any point $x \in G$ 

$$|\tilde c_x - c_x| < 3 \varepsilon.$$

\end{lemma}
Proof.

\begin{enumerate}
\item

Let $s$ be a Jordan curve on $P$ (with possible breaks),
 which intersects $G$ only at a
point $x$, contains inside upstream part $G_x = \{ y \in G| y>x\}$,
and touches $G$ at all vertices $p_i>x$ once (fig. \ref{lemma1.2.eps}). 
Let $f(s)$ be the image of the curve $s$
on the polyhedral surface $\mycap$, and let $g(f(s))$ be the connected
image
of $f(s)$ on $U_G$.

Denote by $u_i$ the image of vertex $p_i$ touched by $g(f(s))$.
Denote by $c'_x= \alpha_x^{-1} \sum_{p_i > x} \alpha_i u_i$ 

Endpoints of the curve $g(f(s))$ are $x'_L$ and $x'_R$.

\begin{figure}[htb]
\begin{center}
\includegraphics[clip]{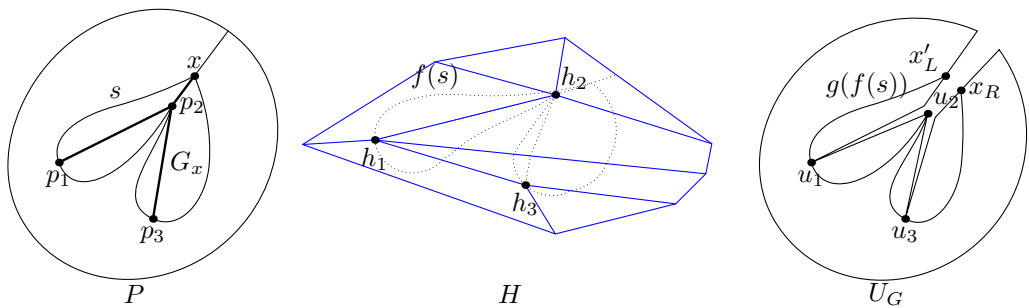} 
\end{center}
\caption{$G_x$ and images of $s$}
\label{lemma1.2.eps}
\end{figure}

\begin{figure}[htb]
\begin{center}
\includegraphics[clip]{lemma1.3.eps} 
\end{center}
\caption{$G'$ and $U'$}
\label{lemma1.3.eps}
\end{figure}
\item 
The outer semi-neighborhood of the curve $s$ has a unique unfolding,
which does not depend from anything beyond this semi-neighborhood.

Let us replace $G_x$ in the graph $G$ by curves $s_i$,
which connect each vertex $p_i$ and $x$ without intersecting
each other
and curve $s$. Denote by $U'$ the unfolding corresponding to
this
constructed curved spanning tree $G'$ (fig. \ref{lemma1.3.eps} b).
 This unfolding is not an $\Lb$-unfolding.

It is possible. Since $s$ is a Jordan curve, it bounds 
a region homeomorphic to a disk [\cite{SJ}]. 

Take the images of the vertex $p_i$ and 
the point $x$ on this disk and connect them by the chord.
Pre-image of this chord is $s_i$.
Chords for different $i$ do not intersect 
each other, so curves $s_i$ do also not intersect each other.

Every vertex $p_i$ has a unique image $u_i$ in the unfolding
$U'$.

\item
Consider the path connecting points $h_i$ and $x$ in the graph $G'$.
There
are two images of $x$ corresponding to this cut:
$x'_{i,L}$ and $x'_{i,R}$ (fig. \ref{lemma1.4.eps} a). 
Vector $ x'_{i,R}-x'_{i,L}$
does not depend on the form of $s_i$  and 
is equal to $M_{rot} \cdot  (x'_{i,c}-u_i) 2 \sin (\beta \alpha_i/2)$,
where
$x'_{i,c}$ is the middle point of the 
segment $x'_{i,l}x'_{i,r}$ (fig. \ref{lemma1.4.eps} b). \\
The distance between $x'_{i,c}$ and $x$ is less than $\varepsilon$.
Denote by $w_i$ the point such that $(w_i - x'_{i,c})= (u_i
- x'_{i,c})
\frac{2 \sin (\beta \alpha_i/2)}{\beta \alpha_i}$. Then $|u_i w_i| <  
\frac{\beta \alpha_i^2}{24}< \beta$ when $\beta \le \varepsilon <
\sqrt{1/5}$.

Let $\tilde u_i$ be given by the formula $\tilde u_i= w_i +
(
x- x_{i,c})$.

Distance $|\tilde u_i -  u_i| < 2\varepsilon$. 

\begin{figure}[htb]
\begin{center}
\includegraphics[clip]{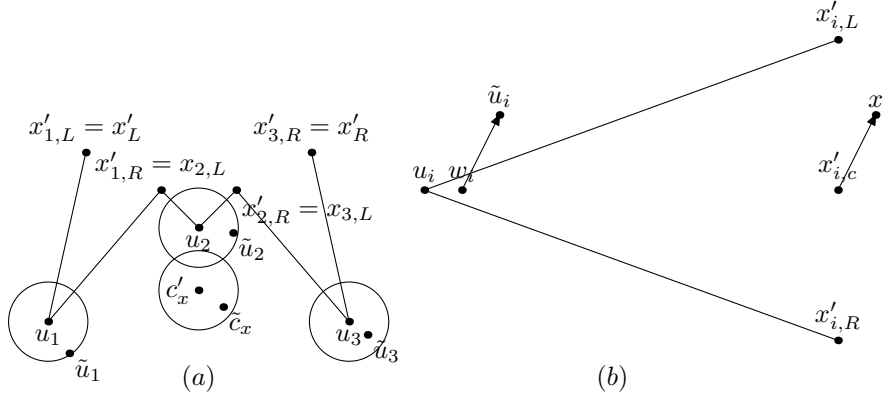} 
\end{center}
\caption{Images of $x$}
\label{lemma1.4.eps}
\end{figure}

Then $x'_{i,R}-x'_{i,L} = M_{rot} \cdot 
(x'_{i,c}-u_i) 2 \sin (\beta \alpha_i/2) = M_{rot} \cdot 
(x'_{i,c}- w_i) \beta \alpha_i = \beta \alpha_i M_{rot} \cdot 
(x- \tilde u_i)$ and 
$|\tilde u_i - u_i| < \varepsilon + \beta \le 2 \varepsilon$.

We get $x'_R- x'_L  = \sum_{\forall i: p_i > x} ( x'_{i,R}-
x'_{i,L}) = \sum_{p_i
> x} \beta \alpha_i M_{rot} \cdot ( x- \tilde u_i) $  

By $(4)$  
$|\tilde c_x = \alpha_x^{-1} \sum_{p_i>x} \alpha_i \tilde u_i$.

As $\tilde c_x$ is a weighted center 
of $\tilde u_i$, and $c'_x$ is a weighted center of $u_i$, we obtain:
$| \tilde c_x - c'_x| < 2 \varepsilon$.

\item
By (3.1) $|u_i - p_i| < \varepsilon$.
Then $|c'_x - c_x | < \varepsilon$ 
and $|\tilde c_x - c_x| < 3 \varepsilon$.

\end{enumerate}

\begin{thm}
For a given infinitesimally curved surface $P$ 
and its partition $L$ there exists an 
$\varepsilon$ such that for the cap $\mycap$ with curvature
$\beta < \varepsilon$, for any non-admissible cut $G$ 
a $\Lb$-unfolding $U_G$ overlaps.

\end{thm}

Proof.

\begin{enumerate}
\item
As $L$ is a non-admissible partition then $0<l(L)<1$.
As $L$ is fixed, denote in this theorem $l(L)$ by $l$.

Consider a cut $G$. 

Let $A$ be a sufficiently small number, 
such that the disk with radius $A$ and center $p_i$ for any $i$
intersects the skeleton $E(L)$ only at the edges 
incident to the point $p_i$.
Let $\gamma$ be the smallest angle between incident 
edges of the skeleton $E(L)$.

Let the unfolding $U_G$ be arranged with
the polygon $P$ satisfying (3.1) and (3.2). 

\item Since the cut $G$ is non-admissible, there exists an edge
$p_i p_j \in G$, $p_i > p_j$ and not satisfying the admissibility 
condition $(2)$: 
$(f_{p_i},p_i p_j) <-l | p_j - p_i| < 0$. 
Let us take a point $a$ on the edge $ p_i p_j$,
such that the distance between $p_i$ 
and this point $a$  is $r_1= \min(A/2,l/2)$. Since
$(a -c_a) = ( (p_i-c_a) + (a-p_i) )$.
Using $| a - p_i| < l/2 < r_1$, we obtain $<a-p_i, p_j-p_i> < l/2 |
p_j - p_i|$.
Then 
$$<a - c_a, p_j - p_i> < -l/2 | p_j - p_i|. \eqno(5)$$

\begin{figure}[htb]
\begin{center}
\includegraphics[clip]{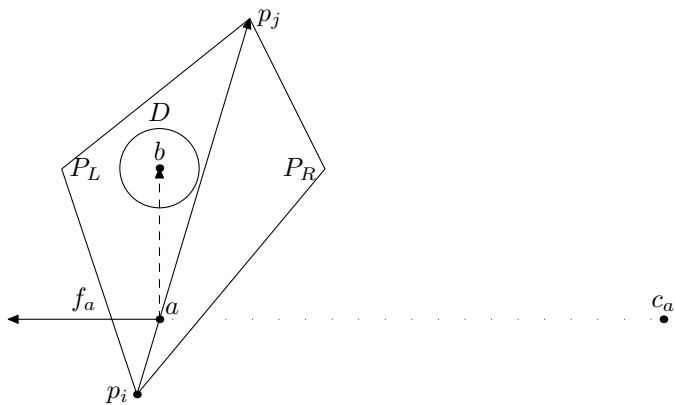} 
\end{center}
\caption{ The point $a$ and the disk $D$.}
\label{thm1}

\end{figure}

\item 
Let show that the unfolding $U_G$ has self-intersections 
in the point $a$, i.e. the right image $a'_R$ 
lies inside interior of $U_G$. 

Denote by $b$ the point $a + \beta M_{rot} \cdot f_a$ (fig. \ref{thm1}).

Denote by $P_L, P_R$ two polygons of the 
partition $L$ incident to the edge $p_i p_j$,
respectively to the left and to the right of the downstream direction.

Let $D$ be a disk with the center $b$ and radius $r_D=\beta \alpha_a
\frac{l}{2}$.

Let us show that $d=|b'- a'_R| < r_D$, 
i.e. $a'_R$ lies inside the image of $P_L$.

\item 
If $\beta$ is small enough ($\beta < \frac{A}{4}$),
then we have $|p_i- a| + |a-b| +r_D \le r_1 + \beta |f_x|+r_D$.

As $f_a \le \alpha_a (c_a -a) \le 1$, we obtain.
$| p_i -a| +|a-b| + r_D \le A/2 + \beta + \beta \frac{l}{2} 
< A/2 + A/4 (1+ l/2) <A$.

The disk $D$ lies inside $P_L \cup P_R$.
Indeed, for any point $x \in D$, 
$\sin (\angle a p_i x) < \frac{r_D+\beta |f_x|}{r_1- \beta | f_x| - r_D} <
\frac{\beta (l/2+1)}{r_1- \beta (l/2+1} < \beta \frac {3}{r_1}$.

For $\beta <  \frac{ \sin(\gamma) r_1}{3}$. $\sin (\angle a p_i x) < \sin(\gamma)$.

Then $ \angle a p_i x < \gamma$ and the  disk $D$ lies inside $P_L \cup P_R$.  

\item
$\sin(\angle x a b ) \le \frac {r_D}{|b-a|} = \frac{\beta \alpha_a l/2}{\beta \alpha_a |a-c_a|} =\frac{ l/2}{|a -c_a|}$.
Using $(5)$ we obtain $\sin(\angle b a p_j)=-\cos(\angle c_a a p_j) >\frac{l/2}{a- c_a}$.
Then the disk $D$ does not intersect with the edge $p_i
p_j$. Hence,
 $ D \subset int(P_L)$.

\item
An image $g(f(b))$ of the point $b$ is the single point $b'$.

An image $g(f(a))$ consists of two points $a'_L, a'_R$. 
Taking into account second condition of Proposition $1$, we obtain
$| (b'- a'_L) - (b-a) | < \varepsilon |(b-a)| =
\varepsilon |f_a| $.

\item

Consider the distance $d = |b'-a'_R|$.

 $d=| b'- a'_R| = | (a'_R - a'_L) - (b'- a'_L)|$.
Using (3.1), 
$b'- a'_L = b-a + \varepsilon |b-a|= b-a + \varepsilon \beta f_a$,
$d < |( a'_R - a'_L) - (b-a) | + \varepsilon \beta |f_a|.$

Using Lemma 1, $a'_R- a'_L=\beta \alpha_a M_{rot} \cdot (a- \tilde c_a)$,
and $|\tilde c_a - c_a|  < 3 \varepsilon$,
$d< \beta \alpha_a | (a- \tilde c_a )  - ( a-  c_a ) | +
\varepsilon \beta  \alpha_a | a - c_a| \le \beta \alpha_a (3 \varepsilon
 + \varepsilon | c_a-a|) < \beta \alpha_a 4 \varepsilon $.

Then when $\beta <
\frac{l}{8}$, we obtain that $d=| a'_R - b'|< r_D$.

Thus, if $\beta$ is small enough, the point $a'_R$ lies inside
the
image of the disk $D$ in the unfolding, i.e. an intersection
has occurred.

\end{enumerate}

\section{Construction of the gadget}

In this section we will construct a gadget $T$ -- a special example
of infinitesimal polygon and its partition.

Our goal is the following theorem:
\begin{thm}
There exists an infinitesimal surface $T$ and its partition
$L_T$ that the following conditions are hold: \\
1. Polygon $T$ is a triangle. \\
2. Partition $L_T$ is non-admissible, i.e. any admissible
graph $G$ of $L_T$ contains non-empty part $G_B$. \\
3. A admissible graph $G$ connects all non-zero weighted vertices
of $T$.  
\end{thm}

Outline of the proof:\\ 
1. The basic idea of a spiral .\\
2.  Constructing  the central ``square'' part of the gadget $T$.\\
3.  Constructing  the whole gadget $T$.\\
4.--7. The proof the theorem  \\

In spite of the fact that the gadget $T$ is very complicated, 
only a few things are needed for the main proof: 
the combinatorial structure of $L_T$, the fact that the partition $L_T$ is convex,
and metric properties, enumerated on fig. \ref{table}.

Proof. 

\begin{enumerate}
\item
At first we will the show basic idea how to construct a partition
with very few admissible spanning trees.

For a fixed vertex $a$ of the cut $G$, the rule $(2)$ defines a
half-plane, which the downstream edge from $a$ should point at.
There
exists at least one such edge, because $L$ is a convex partition.
But
this rule allows this edge not to be unique.
It turns out, that one can construct
special cases for which such continuation is almost unique.

Here we construct a partition, for which, given a starting 
point, we can define any admissible cut unambiguously.

For the center $C$, the number of points in period $n$, the growth
parameter $0 < q \in \mathbb{R}$ and the starting point $s$ (or
endpoint $e$ with index $i_e$) take a logarithmic spiral 
$$f(x) = ( C_x + r_0 q^{x}\cos (\phi_0+ 2\pi x ), C_y + r_0 q^x
\sin(\phi_0 2 \pi x)) \eqno{(6)},$$ 
where $x=i/n$, $r_0,\phi_0$
are parameters to fit $f(0)= s$ (or to fit $f(i_e/n)= e$).

\begin{figure}[htb]
\begin{center}
\includegraphics[clip]{spiral.1.eps} 
\end{center}
\caption{Spiral}
\label{spiral1}
\end{figure}

\begin{figure}[htb]
\begin{center}
\includegraphics[clip]{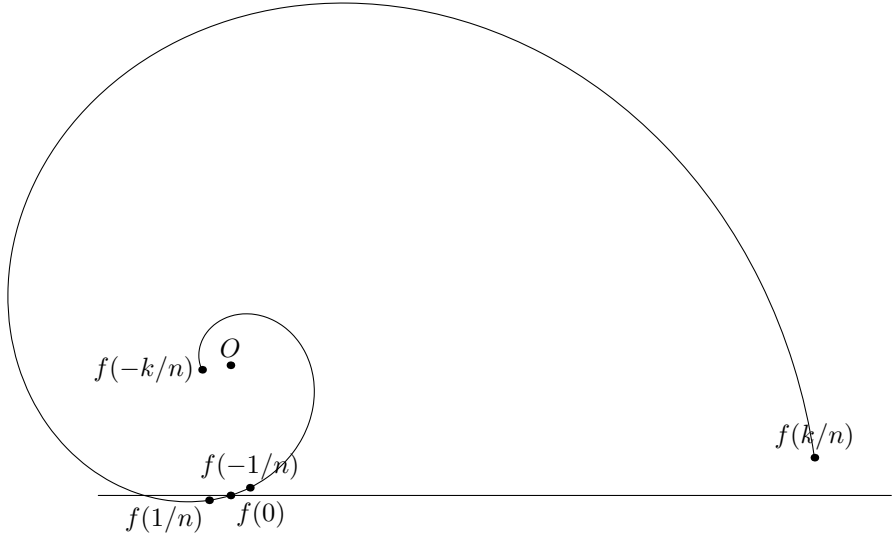} 
\end{center}
\caption{Adjacent points to $f(0)$}
\label{spiral2}
\end{figure}
Let $k,n \in \mathbb{N}$ be an integer numbers 
satisfying following conditions (fig. \ref{spiral1} a): 
$$ \angle O f(0) f(k/n) < \pi/2  \eqno(7.1)$$ 
$$f(0) \in int(\triangle O f(1/n) f(k/n)). \eqno(7.2)$$ 
Let $0<x<1$ be such value that $\angle O f(0) f(x) = \pi/2$.
Then $k/n$ should be less and  close enough to the $x$.
If  $n$ is large enough such $k$ always exists.
Take $n_{max} > n+2k$.

Now take $\mya_0=O$ $\mya_i=f(i/n)$ for $1 \le i \le n_{max}$.
$P_{spiral}=conv(\mya_0,\mya_1,..\mya_{n_{max}-k})$.
Weights $\alpha_0=1$ and $\alpha_i=0$ for all $i >0$.

For any $i<n_{max}$ construct an edge between points $\mya_i$ and
$\mya_{i+1}$.
For $0< i \le k$ construct an edge between points $\mya_0$ and
$\mya_i$.
For $0 <i \le n_{max}-k$ construct an edge between points $\mya_i$
and $\mya_{i+k}$.
This set of edges intersecting with $P_{spiral}$ defines the partition
$L_{spiral}$.
There are $3$ or $4$ angles around any non-bound vertex $\mya_i$.
The biggest angle is equal to $\angle f(1/n) f(0) f(k/n)$. This angle is less than $\pi$ because of $(7.2)$. 
(fig. \ref{spiral1} b).
Thus, all polygons of this partition are convex.

Consider some admissible graph $G$. 

Suppose the vertex $\mya_i$ is downstream to $\mya_0$, 
then $c_{\mya_i}=\mya_0$.

There are some possible ways from $\mya_i$: $O, \mya_{i-1}, \mya_{i+1},
\mya_{i-k},\mya_{i+k}$ (not all of them exist at the same time).

There is only one way with supporting condition $(2)$ : $\mya_i
\mya_{i+1}$ (you see on fig. \ref{spiral2} only $f(1/n)$ is under the line).

Thus, any admissible cut of this partition contains the polyline
$\mya_k, \mya_{k+1}, \ldots, \mya_{n_3}$, where $n_3$ is the
minimal index of the point $p_i$ on the boundary of the convex
hull $P_{spiral}$ ($\mya_{71}, \ldots, \mya_{85}$ on fig. \ref{spiral1} b).

Such unambiguous continuation is the main idea in the construction
of our counterexample.

\medskip

\item Now we construct the gadget $T$ (fig.~\ref{square},~\ref{zoom11},~\ref{zoom8},~\ref{zoom9}).

At first let us construct central part -- the ``square".
This part is centrally symmetrical. 

Two vertices: $\myc_1 = (-70,0); \myc_2= (70,0)$.
These points have equal weights $1/2$. All following vertices
in the gadget have a zero weight.

Add points $\myh_1=(-5,15); \mye_1=(-5,10);$
$\myj_1 =(-3.791,-5.006); \myf_1=(-6.565,21.485); \myf'_1=c_{1,199}=(-6.807,21.404);\mym_1=(-11.002,42.812);
\mym'_1=\myc_{2,139}=(17.124,40.546); \myn_1=(213.886,53.695); \myn'_1=\myc_{1,198}=(-53.695,218.886)$. 
Note that $\myf'_1$ is very close to $\myf_1$ (fig.~\ref{zoom8} b).

Let us in a centrally symmetrical way take points with opposite index.
Add edges between $\myn_1,\myn'_1,\myn_2,\myn'_2$. This is our square.

Then add the following edges 
$\myh_1 \myh_2, \mye_1 \myh_1, \myh_1 \myf_1, \myf_1 \myj_2,
\myf_1 \mym'_1, \myf_1 \myf'_1, \myf'_1 \mym_1, \mye_1 \myf'_1, \mym_1 \mym'_1, \mym'_1 \myn'_1$
and all centrally symmetrical them edges.

Using formula $(6)$, let us make two spirals around $\myc_1,\myc_2$ ending
in $\mye_1,\mye_2$.
Parameters for a spiral: $q=2,n=83,k=70,n_{max}=199$ (we can take
any value of $q$, but if $q$ is small, then $n$ and $k$ must be very
big, if $q$ is bigger -- we can not draw a picture). 
 
Then we make a clock-wise spiral around $\myc_1$ : 
$\myc_{1,0},\myc_{1,1},\ldots, \myc_{1,129}=\mye_1, \ldots \myc_{1,199}$.

Add the following edges: 
From $\myc_1$ to $\myc_{1,i}$ for all $i\le k$.\\
From $\myc_{1,i}$ to $\myc_{1,i+1}$ for all $i<k-1$.\\
From $\myc_{i,i}$ to $\myc_{1,i+k}$ for all $i < k$ (for $i=k$ 
the edge is already added, $\mye_1 \myf'_1$).\\

We have to do something with tails: edges $\myc_{1,i}\myc_{1,i+k}$
where $i>129$.
We change position of the last vertex to the first intersection
of a segment $\myc_{1,i} \myc_{1,i+k}$ (from point $\myc_{1,i}$) with
a line constructed above. 

Now points $\myc_{1,130},\myc_{1,131}$ lie on $\mye_1 \myj_1$, $\myc_{1,132},\ldots
\myc_{1,138}$ lie on $\myj_1 \mym'_2$ ($\mym'_2=\myc_{1,139}$), $\myc_{1,140},\ldots
\myc_{1,150}$ lie on $\mym_2 \myn'_2$,  $\myc_{1,151},\ldots \myc_{1,171}$
lie  on $\myn'_2 \myn_2$,
$\myc_{1,172},\ldots \myc_{1,198}$ lie on $\myn_2 \myn'_1$ ($\myn'_1= \myc_{1,198})$.

We construct second spiral around $\myc_2$ in the same way.
At last, in order to make all angles less than $\pi$, 
ley us  slightly bend all semgnets where ends of tails lie. 
Every ``tail" become shorter and does not change its direction.
 
The ``square" is now slightly concave. 

\begin{figure}[htb]
\begin{center}
\includegraphics[clip]{spiral7.1.eps} 
\end{center}
\caption{Square part of the gadget $T$}
\label{square}
\end{figure}

\begin{figure}[htb]
\begin{center}
\includegraphics[clip]{spiral11.1.eps} 
\end{center}
\caption{Zoom of the square part.}
\label{zoom11}
\end{figure}

\begin{figure}[htb]
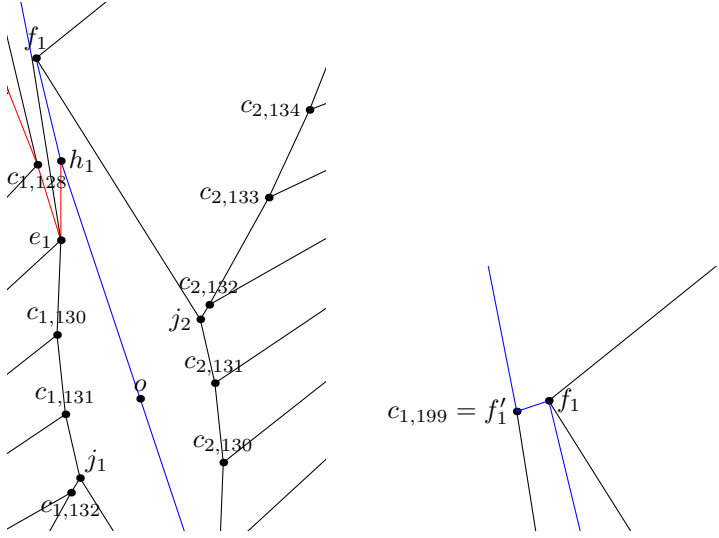

\begin{center}
\begin{tabular}{c c}
\includegraphics[clip]{spiral8.1.eps} & \includegraphics[clip]{spiral10.1.eps}
\\ 
\end{tabular}
\end{center}
\caption{Zoom of the square part and deep zoom around $\myf_1$ with $\myf'_1$}
\label{zoom8}
\end{figure}

\begin{figure}[htb]
\begin{center}
\includegraphics[clip]{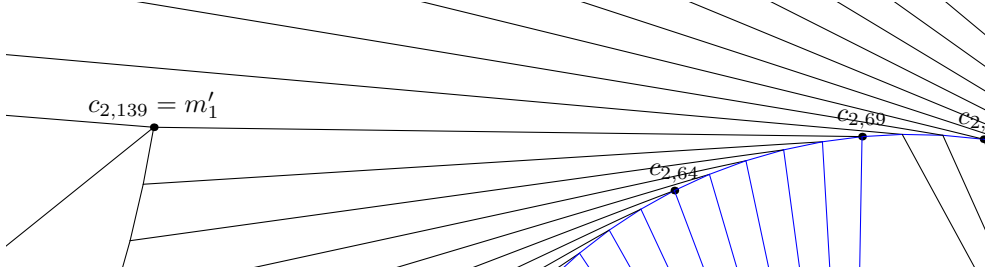} 
\end{center}
\caption{Zoom around the edge $\mym'_1 \myc_{2,69}$}
\label{zoom9}
\end{figure}
\item 
Making a whole gadget (fig.~\ref{gadget}).

Make a triangle with the vertices 
$\myt_i = 7500 (\sin(2 \pi i), \cos(2 \pi i))$, $i \in \{1,2,3\}$.

Make third spiral with center in origin.
Parameters of spiral: $q= 2.4; n=83;k=68; n_{max}=87;$ 
(this spiral works for three centers: $\myc_1,\myc_2,\myo$ so we need
bigger $q$).

Coordinates of starting point $\myo_{0}=(0,-1198.4)$, 
of ending points $\myp=\myo_{87}=(-894.6,-2863.5)$.
Add edges $\myn'_1 \myo_{42}, \myn'_2 \myo_0, \myn_1 \myo_{21}, \myn_2 \myo_{63}$.

Cut tails as in previous case.
Vertices $\myo_{88},\ldots,\myo_{100}$ are on the edge $\myt_1 \myt_2$.
Vertices $\myo_{101},\ldots,\myo_{127}$ are on the edge $\myt_2 \myt_3$.
Vertices $\myo_{128},\ldots,\myo_{153}$ are on the edge $\myt_3 \myt_1$.
Bend edges $\myt_1 \myt_2, \myt_2 \myt_3, \myt_3 \myt_1$ 
to makes all angles strictly less than $\pi$. 
These edges became slightly concave polylines.

The gadget is ready. The polygon $T$ is $\triangle  \myt_1 \myt_2 \myt_3$.

Partition $L_T$ is defined as set of the vertices and edges.

To prove that any polygon of $L_T$ is convex we need to investigate
every vertex and find that any incident angle is less than $\pi$.

For the vertices on the spiral it was made by construction, also for
the vertices on the spiral tails. 
Other points $\mye_i,\myh_i,\myf_i,\myf'_i,\myj_i,\mym_i,\mym'_i,\myn_i,\myn'_i$ can be
investigate manually.
Also these conditions were verified by the computer program.
\begin{figure}[htb]
\begin{center}
\includegraphics[clip]{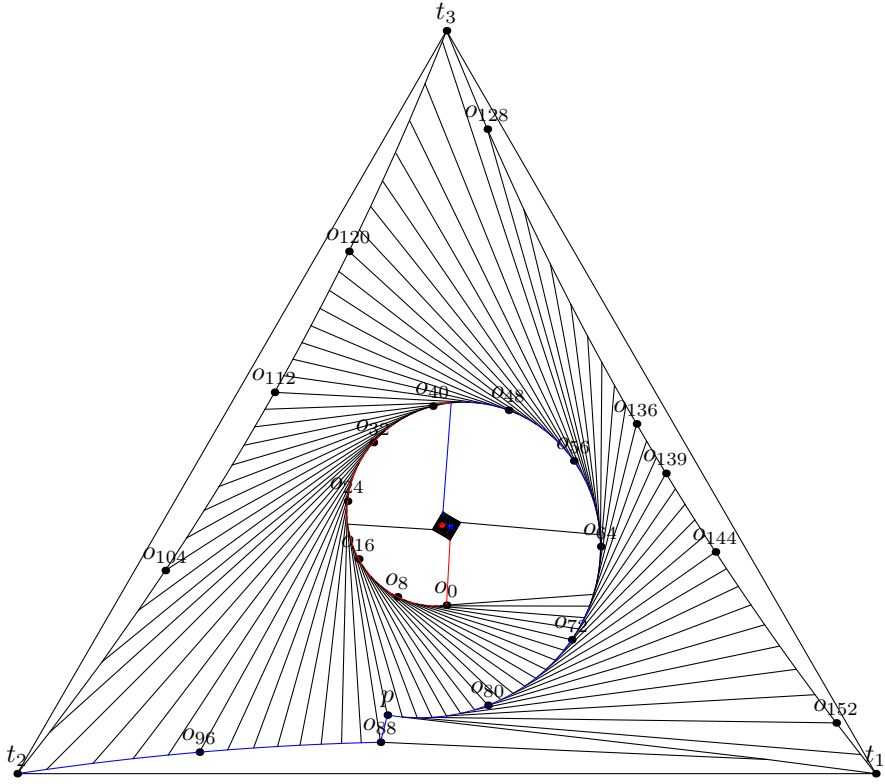} 
\end{center}
\caption{The gadget $T$ (in the center the square part of $T$)}
\label{gadget}
\end{figure}

Coordinates of the gadget and the program one can downloaded 
from the web-page
http://dcs.isa.ru/taras/durer and explore it in detail.

Further  we will need metric properties enumerated on fig.\~ref{table}. 
This table means that $\myy$ is only one vertex adjacent to $\myx$
that $\angle \myy \myx c_\myx$ is obtuse. For all other vertices 
corresponding angle is acute.

\begin{figure}[h]
\begin{tabular}{ ||l | l| l  ||}
\hline
Vertex ($\myx$) & Rotation center ($c_\myx$)  & Possible exit  $\myy$ 
\\ 
$\myc_{i,j}$ $(j<91, j \ne 70)$  & $\myc_1$  & $\myc_{i,j+1}$ \\
$\mye_i$  & $\myc_i$  & $\myh_i$ \\
$\myh_i$  & $\myc_i$  & $\myh_{3-i}$ \\
$\myh_{i}$  & $\myc_{3-i}$  & $\myf_{i}$ \\
$\myf_{i}$  & $\myc_{3-i}$  & $\myg_{i}$ \\ 
$\myg_{i}$  & $\myc_{3-i}$  & $\mym_{i}$ \\
$\mym_{i}$  & $\myc_{3-i}$  & $\myc_{1,81}$ \\
$\myf'_{i}$  & $\myc_{3-i}$  & $\mym_i$ \\
\hline
$\myh_i$  & $\myo$  & $\myf_{i}$ \\
$\myf_i$  & $\myo$  & $\myc_{i,80} (\mym'_{3-i})$ \\
$\myc_{i,80} (\mym'_{3-i})$  & $\myc_i$  & $\myc_{i,10}$ \\
\hline
$\myc_{1,91}=\myn'_1$  & $\myc_1,\myc_2,\myo$  & $\myo_{42}$ \\
$\myc_{2,91}=\myn'_2$  & $\myc_1,\myc_2,\myo$  & $\myo_{0}$ \\
$\myo_{j}$ ($j \le 99$)  & $\myc_1,\myc_2,\myo$  & $\myo_{j+1}$ \\ 
$\myo_{100}$  & $\myc_1,\myc_2,\myo$  & $\myt_2$ \\
\hline
\end{tabular}
\caption{Necessary metric conditions of $T$}
\label{table}
\end{figure}

These conditions were verified by a simple computer program and
can be confirmed manually as well.

 \item

Divide $G_A$ into $4$ graphs $G_1, G_2, G_{1,2},G_0$ that do
not
share edges. $G_1 = \{ e \in G | \myc_1 > e {\rm\; and\; not\;}
(\myc_2
> e) \}$, $G_2 = \{ e \in G | \myc_2 > e {\rm\; and\; not\;} (\myc_1
>
e) \}$, $G_1 = \{ e \in G | \myc_1 > e {\rm\; and\;} \myc_2 > e \}$
and $G_0$ all the rest.

We adopt the convention that a graph containing 
an edge also contains its vertices.


\item Consider the first case $G_{1,2}=\emptyset$.

Then the $G_1,G_2$ polylines separately connecting  with $\partial
T$ or $G_B$.

Using the
properties of the gadget $T$ (fig.~\ref{table}) and 
the condition $(2)$, we can define the edges
of $G_1$ and $G_2$ as follows: 

The graph $G_1$ starts from the point $c_1$ and goes by
any possible
edge to the point $\myc_{1,i}$ ($i < k)$ of the spiral. Further $G_1$ continued
unambiguously along the follwing path: $\myc_1,\myc_{1,i},\myc_{1,i+1},
\ldots \myc_{1,69},\ldots,\myc_{1,129}=\mye_1, 
\myh_1, \myh_2,\myf_2,\myf'_2,\mym_2,\myc_{1,140},\myc_{1,141},\ldots,
\myc_{1,150},\myc_{2,198}=\myn'_2, \myo_0, \myo_1, \ldots, 
\myo_{87}=\myp,\myo_{88},\ldots,\myo_{99},\myt_2$.

The graph $G_1$ is a part of this path from $\myc_1$ to the
first point of $G_B$ or $t_2$.

We can define the path for $G_2$ similarly:
$\myc_2,\myc_{2,i},\myc_{2,i+1},
\ldots \myc_{2,69},\ldots, \myc_{1,129}=\mye_2, \myh_2, \myh_1,\myf_1,
\myf'_1,\mym_1,\myc_{2,140},\myc_{2,141},\ldots,
\myc_{2,150},\myc_{1,198}=\myn'_1, 
\myo_{42},\myo_{43},\ldots, \myo_{87}=
\myp,\myo_{88},\ldots,\myo_{99},\myt_2$.
Again $G_2$ is a part of this path from $\myc_2$ to the first point
in $G_B$ or $\myt_2$.

We see that if $G_B = \emptyset$, both $G_1$ and $G_2$ cannot
avoid the bridge $\myh_1 \myh_2$ and intersects each other. Then if $G_{1,2}$ is
empty than $G_B$ is not empty.

\item
Consider the case, that $G_1$ and $G_2$ are in different arc
components of $G$. 

As $P_T$ has only $3$ vertices, and every component of $G_B$
should have at least $2$ endpoints in $P_T$, $G_B$ is a connected
graph.
Then for some $i \in [1,2]$ the graph $G_i$ falls into $\partial
T$ and $G_{3-i}$ falls into $G_B$.
Because $G_i$ falls into $\myt_2$, $G_B$ is a polyline connecting
vertices $\myt_1$ and $\myt_3$. 
But this polyline cannot reach ``square" part of $T$ without
intersection with $\myo_{61},\ldots,\myo_{85}$, the part of $G_i$.

Then if $G_{1,2} = \emptyset$, $G_1$ and $G_2$ falls separately
into $G_B$.
Hence, $G$ connects vertices $\myc_1,\myc_2$. 
We have proved theorem for $G_{1,2} = \emptyset$.

\item Consider the case $G_{1,2} \ne \emptyset$.
$G_1$ and $G_2$ go through the bridge $\myh_1 \myh_2$. $G_2$ does not
intersect $G_1$ 
above $\myh_1$, and $G_1$ does not intersect $G_2$ above $\myh_2$.

Then $G_{1,2}$ can start only from one $\myh_i$, $i \in \{1,2\}$.

We can unambiguously continue the path $G_{1,2}$ from each 
of these points until it falls into $G_B$ or $\partial T$. 
For every $x \in G_{1,2}$ the rotation center $c_x$ is the origin $\myo$, midpoint between
$\myc_1$ and $\myc_2$.

From the vertex $\myh_i$ the graph $G_{1,2}$ continued unambiguously:
 $\myh_i, \myf_i, \mym'_i, \myo_{3-i,69}$.
Note that $\myo_{i,69} \in G_i$ ($\myo_{i,69}$ is a last point which
has common edge with $\myc_i$). $G_{1,2}$ cannot fall into $G_i$,
so it falls into $G_B$ in one of the points $\myf_1,\myf_2,\mym'_1,\mym'_2$.

As $G_{1,2} \ne \emptyset$, it is clear that $G$ 
connects two vertices $\myc_1,\myc_2$, because $G_1,G_2,G_{1,2}$ 
have a common point -- the outfall of $G_1,G_2$ and origin of $G_{1,2}$.

Thus, if $G_{1,2} \ne \emptyset$, then $G_B \ne \emptyset$ and
$G$ connects $\myc_1, \myc_2$.

This completes the proof of the Theorem 2.

\end{enumerate}

\section{Proof of the main theorem}

\begin{figure}[htb]
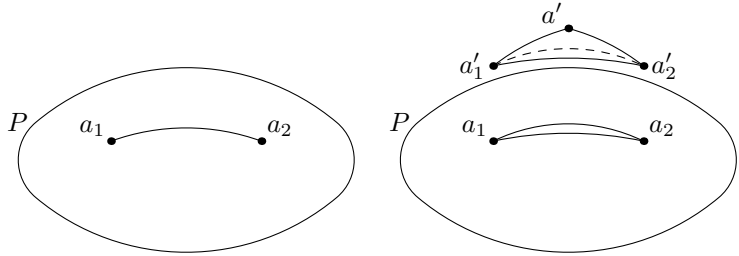

\begin{center}
\begin{tabular} {c c}
\includegraphics[clip]{coneshaping.1.eps} & 
\includegraphics[clip]{coneshaping.2.eps}
\\ 
\end{tabular}
\end{center}
\caption{Cone-shaping}
\label{coneshaping}
\end{figure}

{\bf Definition 19. }
Let $Q$ be a polyhedral surface.
Let $h_1,h_2$ be two vertices satisfying $curv(h_1)+curv(h_2)
< 2 \pi$.

Take a geodesic $h_1h_2$. Cut the surface $Q$ by this geodesic.
Take the double triangle $a'_1a'_2a'$ with $a'_1a'_2=h_1h_2;
\angle a'a'_1a'_2= curv(h_1)/2; \angle a'a'_2a'_1= curv(h_2)/2
$.
Cut this triangle by $u_1u_2$. Glue the triangle and the surface
by the lines of the cut, so that the new glued surface $Q'$ is
homeomorphic to sphere (disk).

The surface $Q'$ is intrinsically convex and satisfies the conditions
of 
Alexandrov's existence theorem.

Hence $Q'$ is the surface of a convex polyhedron (or part of
the surface). This polyhedral surface
 contains one vertex $v$
with curvature $curv(v)=curv(p_1)+curv(p_2)$
instead of the two vertices $p_1,p_2$.

This procedure is called {\it cone-shaping}.

\begin{lemma}
Let $Q$ be the surface of a polyhedron, let $L_Q$ be a convex geodesic partition.
For a given vertex $v$ one can modify $Q$ by changing 
the neighborhood of $v$ into a scaled copy of the gadget cap $T_v$ with curvature $curv(v)$. 
The obtained polyhedral surface $Q'$ has a corresponding convex geodesic
partition $L'$.
\label{lemma2}
\end{lemma}
Proof.

\begin{enumerate} 

\item 
Denote by $e_i$ the edges incident to $v$ (fig.~\ref{newpartition} a).
Let $\gamma$ be the biggest angle of the partition $L$ around
the vertex $v$.
Let $r$ be the length of the smallest edge $e_i$. 
Put a point $z_i$ on each edge $e_i$ on distance $r/3$ from $v$.

For adjacent $e_i e_j$ add an edge between $z_i z_j$.
Denote by $Z$ the polyhedral surface bounded by the constructed cycle
$z_i$ and containing $v$.

Let us scale  $T_v$ to make diameter $d(T_v) <  r/3 \cos(\gamma/2)$.

\begin{figure}[htb]
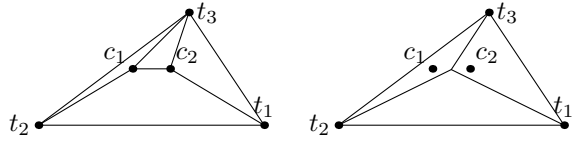

\begin{center}
\begin{tabular} {c c}
\includegraphics[clip]{coneshapedT.1.eps} &  \includegraphics[clip]{coneshapedT.2.eps}
\\ 
\end{tabular}
\end{center}
\caption{Schematic $T_v$ with real edges and cone-shaped
\label{coneshapedT}
$T'$}
\end{figure}
Let $T'$ be a cone-shaped surface of  $T_v$ (fig.~\ref{coneshapedT}).
Then $d(T') < 2 d(T_v)$ for small enough $curv(v)$.
The polyhedral surface $T'$ has one vertex with curvature $curv(v)$.
Then $T'_v$
is isometric to a fragment of the surface of $Q$ around the vertex
$v$.

\item Cut from $Q$ the piece isometric to $T'$ and replace it
into $T_v$. 
This gluing is made isometrically, 
because $T'$ and $T_v$ has an isometric neighborhood
of the border.

$Q'$ is a new surface.
Surface $Q'$ is an intrinsically convex.
Then it is a surface of some convex polyhedron. 

\begin{figure}[htb]
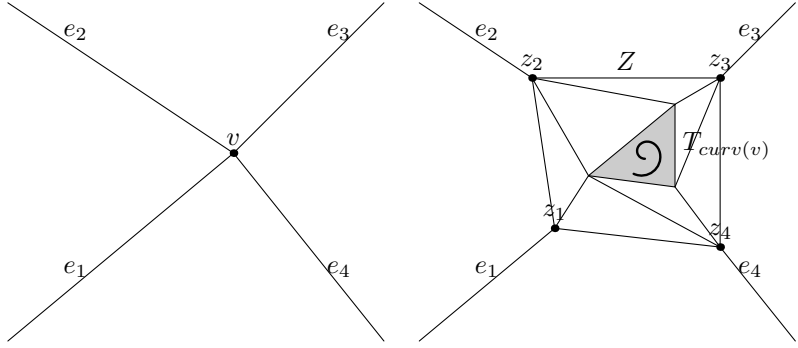

\begin{center}
\begin{tabular} {c c}
\includegraphics[clip]{newpartition.1.eps} &  \includegraphics[clip]{newpartition.2.eps}
\\ 
\end{tabular}
\end{center}
\caption{Replacing neighborhood of $h_i$ into $T_i$}
\label{newpartition}
\end{figure}
\item 
Let us define a partition $L_{Q'}$ for $Q'$ (fig.~\ref{newpartition} b). 
Inside $T_i$ take the partition $L_v$. 

Outside $Z$ the partition $L_{Q'}$  equals to $L_Q$.
On $T_i$ the partition $L_{Q'}$ equals to $L_i$.
The region outside $T_i$ and inside $Z$ should be somehow triangulated.

It is clear that $L_{Q'}$ is a geodesic convex partition.

Additional points $z_i$ only needed to make sure that all polygons 
of the partition are convex. 
\end{enumerate}

\begin{thm} There exists a surface of a convex
polyhedral surface $P$ and its geodesic convex partition $L_P$ 
 such that there are no connected non-overlapping
$L$-unfoldings of $P$. 
\end{thm}

Proof. 

\begin{enumerate}

\item 

The technique of construction a counterexample is similar to
the technique in another counterexamples without 
non-overlapping unfolding (\cite{GlazTar08},\cite{Tar99}): 
we take a polyhedron $Q$ and replace its vertex with a special gadget.
Any spanning tree of the modified polyhedron $Q'$ 
has a corresponding spanning tree in the original polyhedron $Q$
(every time we have to show that such correspondence exists).

Every spanning tree in $Q$ has at least one end. 
The unfoding of $Q'$ has overlappings in a gadget 
corresponding to this end, because of the properties of the gadget.

\item
By Theorem 1, for the gadget $T$, there exists an $\varepsilon$
such that
the cap $T_{\beta}$ is not admissible for $\beta < \varepsilon$.
Let $Q$ be a convex polyhedral surface  with the minimal distance between
vertices at least $2 d(T)$ and the curvature of any
vertex less than $\varepsilon$.

For example, 
the prism based on the regular $N$-gon, where $N> \frac{2 \pi
}{ \varepsilon}$, satisfies
this condition.

Using Lemma~\ref{lemma2}, change every vertex $h_i$  of $Q$ into a corresponding
gadget  $T_i$ with the curvature $curv(h_i)$.

We obtain a surface $P$ and its partition $L_P$.
Denote by $L_i$ the part of $L_P$ corresponding to the gadget $T_i$.

Every vertex $w$ of $P$ is a vertex $c_1$  or $c_2$ of
$L_i$ by construction. 
\item 

Let $G$ be a spanning tree of the polyhedral surface $P$ such that $G
\subset L_P$
and the corresponding unfolding $U_G$ does not have any intersections.
Denote by $G'$ the graph consisting of edges $e \in G$ which
divide $G$
into two parts $G_1,G_2$, where each part contains vertices of
non-zero curvature.
It is easy to see that we can unfold $S$ by cutting only $G'$.
Not cutted vertices have a zero curvature.  The obtained unfolding
$U_{G'}$ is equal to $U_G$.

Consider a fragment $L_i$, denote by $F_i = G' \cap L_i$.!!
As $U_{G'}$ is a non-overlapping unfolding, $F_i$ is a cut. 

By the second condition of Theorem 2, $F_i$ has a non-empty subgraph
$F_{i,B}$ (which consists of $B$-edges of $F_i$).

Therefore $F_i$ is connected to $G'\setminus F_i$ by at least
two points.
By the third condition of Theorem 2, the graph $F_i$ is connected.
 
Since $F_{i,B}\subset G'$, each endpoint of $F_{i,B}$ should
be the connection between $F_i$ and $G' \setminus F_i$.
Then $F_i$ and $G' \setminus F_i$ have at least $2$ connections.

Replace every fragment $F_i$ in $G'$ into a single vertex.
We obtain a graph $G''$.
This graph $G''$ has a cycle, because the degree of any its vertex
is at least $2$.

Since any $F_i$ is connected, $G'$ also contains a cycle.

\item

Note that there exists a polyhedral surface $P'$ without zero-curvature
vertices,
combinatorially equivalent and arbitrary near to $P$.

Let show this.
Let any vertex $h_i$ with zero-curvature be lying inside 
some face $f$. If some $h_i$ lies on the edge of $P$, 
we can slighlty change the position of $h_i$ 
without changing the combinatorial structure and properties of $P$.

For each vertex $h_i$ of the partition $L$ which is not
a real vertex of $P$ 
take $h'_i = h_i + \varepsilon (d^2-r_i^2) n_f$, 
where $d$ is the diameter of $P$ (as a body), $r_i$ is a distance
from $h_i$ to the center of the real facet $f$ of $P$ where 
this point is lying, $n_f$ is a normal vector of the facet $f$. 
Then $P'=conv(P,conv_i(h'_i))$.
If $\varepsilon$ is small enough all edges and vertices
of $P$ exist in $P'$.

Then $P'=P \cup P_{f_1} \cup P_{f_2} \ldots P_{f_n}$, where
$P_{f_j}$ is a $conv(f_j,conh_{h_i \in f_j}(v'_i))$ for each
face
$f_j$.  
Since all vertices $h'_i$ of some $P(f_j)$ are lying on a paraboloid,
any vertex $h'_i$ cannot lie in the interior of $P_{f_j}$ and
have a non-zero curvature.
Then every vertex of $P'$ has a non-zero curvature.

Also, for a small enough $\varepsilon$ the surface $P'$ has the
partition  $L_{P'}$
corresponding to $L_P$ without any combinatorial changes.

The number of subtrees of $L_P$ is finite, so for a small enough
$\varepsilon$ 
any $L_{P'}$-unfolding of $P'$ overlaps as the corresponding
unfolding in $P'$.

Thus, $P'$ is also a counterexample.

\end{enumerate}

\section{A simpler counterexample}

\begin{figure}[htb]
\begin{center}
\includegraphics[clip]{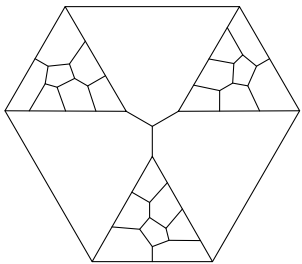} 
\caption {Tutte graph $G_T$}
\label{tutte}
\end{center}
\end{figure}

Let $P_{needle}$ be a polyhedron which holds the following conditions:
\\
1. The edge-skeleton of $P_{needle}$ is isomorphic to the Tutte graph
(fig. \ref{tutte}) 
\cite{Tutte46} (this graph does not have a Hamiltonian cycle).\\
2. $P_{needle}$ has $2$ adjacent vertices $h_1,h_2$ with the sum
of curvature close to $4 \pi$. \\

Such polyhedron exists. 
It can be constructed in the following way.

Using the Steinitz-Rademacher theorem \cite{SteinRad34}, 
take the polyhedron $P$ with its skeleton dual to the Tutte graph $G_T$

Take some edge $e$. 
Let the polyhedron $P^{*}$ be polar to $P$ with the center $O$
of
the polar transformation close to the center of the edge $e$.

The obtained $P^{*}$ has two vertices $h_1,h_2$ incident
to the edge $e^*$. If $O$ is close enough to the edge $e$, 
the projection of every vertex $h_i, i>2$ into 
line $h_1h_2$ lies inside the edge $e^*$.
Squeeze affine $P^{*}$ in two directions 
perpendicular to $e^*$ to make $P^*$ lie in the neighborhood of $e^*$.  

Then the combinatorial structure of $E(P^{*})$ is equal to $G_T$ and 
$P^*$  has 
two adjacent vertices $h_1,h_2$ with the sum of the
curvatures near to $4\pi$. The total curvature of all other
vertices is arbitrary small.

Using Lemma~\ref{lemma2}, replace any vertex of $P^{*}$ except $h_1,h_2$
into the gadget. 
We obtain $P_{needle}$.

Repeat the argument from Theorem 2: let $G$ be any
spanning tree of $P_{needle}$ and $G'$ its non-zero curvature
part. 

Any part of $G'$ inside each $Z_i$ is connected. 
Indeed, any component $F$ of $G' \cap Z_i$ can have a vertex 
with degree $1$ only in $z_i$ and $c_{1,i}, c_{2,i}$. 
If this component $F$ contains non-zero curvature vertices, 
then by third condition of theorem 2 it has two additional ends.

Any component of $F$ of $G' \cap Z_i$ should contain at least
two vertices from $z_{i,1},z_{i,2},z_{i,3}$. 
Hence, $G' \cap Z_i$ is connected graph.

For each $i$ replace $G' \cap Z_i$ into single vertex. 
The obtained graph $G''$ has degree $1$ only in vertices
$h_1,h_2$. As $G'$ does no have cycles, $G''$ is a
Hamiltonian path from $h_1$ to $h_2$. 
Adding to $G'$ theedge $h_1 h_2$ we obtain a Hamiltonian cycle 
in $G_T$.  This contradiction proves the statement.
 
$P_{needle}$ contains 43 gadgets and $2+ 43 \cdot 442=19008$
vertices.

\end{document}